\documentstyle[12pt,epsf]{article}

  \def\pp{{\mathchoice
          {
              \kern 1pt%
              \raise 1pt
              \vbox{\hrule width5pt height0.4pt depth0pt
                    \kern -2pt
                    \hbox{\kern 2.3pt
                          \vrule width0.4pt height6pt depth0pt
                          }
                    \kern -2pt
                    \hrule width5pt height0.4pt depth0pt}%
                    \kern 1pt
           }
            {
              \kern 1pt%
              \raise 1pt
              \vbox{\hrule width4.3pt height0.4pt depth0pt
                    \kern -1.8pt
                    \hbox{\kern 1.95pt
                          \vrule width0.4pt height5.4pt depth0pt
                          }
                    \kern -1.8pt
                    \hrule width4.3pt height0.4pt depth0pt}%
                    \kern 1pt
            }
            {
              \kern 0.5pt%
              \raise 1pt
              \vbox{\hrule width4.0pt height0.3pt depth0pt
                    \kern -1.9pt  
                    \hbox{\kern 1.85pt
                          \vrule width0.3pt height5.7pt depth0pt
                          }
                    \kern -1.9pt
                    \hrule width4.0pt height0.3pt depth0pt}%
                    \kern 0.5pt
            }
            {
              \kern 0.5pt%
              \raise 1pt
              \vbox{\hrule width3.6pt height0.3pt depth0pt
                    \kern -1.5pt
                    \hbox{\kern 1.65pt
                          \vrule width0.3pt height4.5pt depth0pt
                          }
                    \kern -1.5pt
                    \hrule width3.6pt height0.3pt depth0pt}%
                    \kern 0.5pt
            }
        }}

  \def\mm{{\mathchoice
                       {
                             \kern 1pt
               \raise 1pt    \vbox{\hrule width5pt height0.4pt depth0pt
                                  \kern 2pt
                                  \hrule width5pt height0.4pt depth0pt}
                             \kern 1pt}
                       {
                            \kern 1pt
               \raise 1pt \vbox{\hrule width4.3pt height0.4pt depth0pt
                                  \kern 1.8pt
                                  \hrule width4.3pt height0.4pt depth0pt}
                             \kern 1pt}
                       {
                            \kern 0.5pt
               \raise 1pt
                            \vbox{\hrule width4.0pt height0.3pt depth0pt
                                  \kern 1.9pt
                                  \hrule width4.0pt height0.3pt depth0pt}
                            \kern 1pt}
                       {
                           \kern 0.5pt
             \raise 1pt  \vbox{\hrule width3.6pt height0.3pt depth0pt
                                  \kern 1.5pt
                                  \hrule width3.6pt height0.3pt depth0pt}
                           \kern 0.5pt}
                       }}

\catcode`@=11
\def\un#1{\relax\ifmmode\@@underline#1\else
        $\@@underline{\hbox{#1}}$\relax\fi}
\catcode`@=12

\let\du=\du                     

\def\a{\alpha}
\def\b{\beta}

\def\d{\delta}

\def\g{\gamma}

\def\j{\psi}

\def\l{\lambda}

\def\o{\omega}
\def\p{\pi}
\def\q{\theta}
\def\r{\rho}
\def\s{\sigma}
\def\t{\tau}

\def\x{\xi}
\def\z{\zeta}
\def\D{\Delta}
\def\F{\Phi}

\def\J{\Psi}
\def\L{\Lambda}

\def\S{\Sigma}

\def\ve{\varepsilon}
\def\vf{\varphi}

\def\vq{\vartheta}

\def\cf{{\cal F}}

\def\ck{{\cal K}}
\def\cl{{\cal L}}

\def\cz{{\cal Z}}

\def\bo{{\raise-.5ex\hbox{\large$\Box$}}}               
\def\pa{\partial}                                       
\def\de{\nabla}                                         
\def\TH{{\raise.2ex\hbox{$\displaystyle \bigodot$}\mskip-4.7mu \llap H \;}}
\def\face{{\raise.2ex\hbox{$\displaystyle \bigodot$}\mskip-2.2mu \llap {$\ddot
        \smile$}}}                                      

\def\sp#1{{}^{#1}}                              
\def\Bar#1{\overline{#1}}                       
\def\sbar#1{\stackrel{*}{\Bar{#1}}}             

\def\VEV#1{\left\langle #1\right\rangle}        
\def\abs#1{\left| #1\right|}                    
\def\leftrightarrowfill{$\mathsurround=0pt \mathord\leftarrow \mkern-6mu
        \cleaders\hbox{$\mkern-2mu \mathord- \mkern-2mu$}\hfill
        \mkern-6mu \mathord\rightarrow$}
\def\dvec#1{\vbox{\ialign{##\crcr
        \leftrightarrowfill\crcr\noalign{\kern-1pt\nointerlineskip}
        $\hfil\displaystyle{#1}\hfil$\crcr}}}           
\def\dt#1{{\buildrel {\hbox{\LARGE .}} \over {#1}}}     

\def\frac#1#2{{\textstyle{#1\over\vphantom2\smash{\raise.20ex
        \hbox{$\scriptstyle{#2}$}}}}}                   
\def\sfrac#1#2{{\vphantom1\smash{\lower.5ex\hbox{\small$#1$}}\over
        \vphantom1\smash{\raise.4ex\hbox{\small$#2$}}}} 
\def\bfrac#1#2{{\vphantom1\smash{\lower.5ex\hbox{$#1$}}\over
        \vphantom1\smash{\raise.3ex\hbox{$#2$}}}}       
\def\afrac#1#2{{\vphantom1\smash{\lower.5ex\hbox{$#1$}}\over#2}}    

\def\[{\lfloor{\hskip 0.35pt}\!\!\!\lceil}
\def\]{\rfloor{\hskip 0.35pt}\!\!\!\rceil}

\def\du#1#2{_{#1}{}^{#2}}

\def\fracm#1#2{\hbox{\large{${\frac{{#1}}{{#2}}}$}}}
\def\ha{{\fracmm12}}

\def\un{\underline}
\def\fracmm#1#2{{{#1}\over{#2}}}

\def\low#1{{\raise -3pt\hbox{${\hskip 0.75pt}\!_{#1}$}}}

\def\Dot#1{\buildrel{_{_{\hskip 0.01in}\bullet}}\over{#1}}
\def\dt#1{\Dot{#1}}

\newskip\humongous \humongous=0pt plus 1000pt minus 1000pt
\def\caja{\mathsurround=0pt}
\def\eqalign#1{\,\vcenter{\openup2\jot \caja
        \ialign{\strut \hfil$\displaystyle{##}$&$
        \displaystyle{{}##}$\hfil\crcr#1\crcr}}\,}
\newif\ifdtup

\def\ref#1{$\sp{#1)}$}

\def\np#1#2#3{Nucl.~Phys.~{\bf B{#1}} (19{#2}) #3}

\def\cqg#1#2#3{Class.~and Quantum Grav.~{\bf {#1}} (19{#2}) #3}
\def\cmp#1#2#3{Commun.~Math.~Phys.~{\bf {#1}} (19{#2}) #3}

\def\ijmp#1#2#3{Int.~J.~Mod.~Phys.~{\bf A{#1}} (19{#2}) #3}

\def\ibid#1#2#3{{\it ibid.}~{\bf {#1}} (19{#2}) #3}

\parskip=\medskipamount                 
\thicklines                         

\begin{document}


\thispagestyle{empty}               

\def\border{                                            
        \setlength{\unitlength}{1mm}
        \newcount\xco
        \newcount\yco
        \xco=-24
        \yco=12
        \begin{picture}(140,0)
        \put(-20,11){\tiny Institut f\"ur Theoretische Physik Universit\"at
Hannover~~ Institut f\"ur Theoretische Physik Universit\"at Hannover~~
Institut f\"ur Theoretische Physik Hannover}
        \put(-20,-241.5){\tiny Institut f\"ur Theoretische Physik Universit\"at
Hannover~~ Institut f\"ur Theoretische Physik Universit\"at Hannover~~
Institut f\"ur Theoretische Physik Hannover}
        \end{picture}
        \par\vskip-8mm}

\def\headpic{                                           
        \indent
        \setlength{\unitlength}{.8mm}
        \thinlines
        \par
        \begin{picture}(29,16)
        \put(75,16){\line(1,0){4}}
        \put(80,16){\line(1,0){4}}
        \put(85,16){\line(1,0){4}}
        \put(92,16){\line(1,0){4}}

        \put(85,0){\line(1,0){4}}
        \put(89,8){\line(1,0){3}}
        \put(92,0){\line(1,0){4}}

        \put(85,0){\line(0,1){16}}
        \put(96,0){\line(0,1){16}}
        \put(92,16){\line(1,0){4}}

        \put(85,0){\line(1,0){4}}
        \put(89,8){\line(1,0){3}}
        \put(92,0){\line(1,0){4}}

        \put(85,0){\line(0,1){16}}
        \put(96,0){\line(0,1){16}}
        \put(79,0){\line(0,1){16}}
        \put(80,0){\line(0,1){16}}
        \put(89,0){\line(0,1){16}}
        \put(92,0){\line(0,1){16}}
        \put(79,16){\oval(8,32)[bl]}
        \put(80,16){\oval(8,32)[br]}

        \end{picture}
        \par\vskip-6.5mm
        \thicklines}

\border\headpic {\hbox to\hsize{
\vbox{\noindent DESY 99 -- 050  \hfill revised version \\
ITP--UH--10/99 \hfill hep-th/9904196   }}}

\noindent
\vskip1.3cm
\begin{center}

{\Large\bf Exact Hypermultiplet Dynamics}
\vglue.1in
{\Large\bf in Four Dimensions}
\vglue.3in

Sergei V. Ketov~\footnote{Supported in part 
by the `Deutsche Forschungsgemeinschaft'}

{\it Institut f\"ur Theoretische Physik, Universit\"at Hannover}\\
{\it Appelstra\ss{}e 2, 30167 Hannover, Germany}~\footnote{Also at 
High Current Electronics Institute of the Russian Academy of Sciences, 
Siberian Branch,\newline
${~~~~~}$  Akademichesky~4, Tomsk 634055, Russia}
\\
{\sl ketov@itp.uni-hannover.de}
\end{center}
\vglue.2in
\begin{center}
{\large\bf Abstract}
\end{center}

\noindent
We use N=2 harmonic and projective superspaces to formulate the most 
general `Ansatz' for the $SU(2)_R$-invariant hypermultiplet low-energy 
effective action (LEEA) in four dimensions, which describes the 
two-parametric family of the hyper-K\"ahler metrics generalizing the
Atiyah-Hitchin metric. We then demonstrate in the very explicit and
manifestly N=2 supersymmetric way that the (magnetically charged, massive) 
single hypermultiplet LEEA in the underlying non-abelian N=2 supersymmetric 
quantum field theory can receive both perturbative (e.g., in the
Coulomb branch) and non-perturbative (e.g., in the Higgs branch)
quantum corrections. The manifestly N=2 supersymmetric Feynman rules
in harmonic superspace can be used to calculate the perturbative 
corrections described by the Taub-NUT metric. The non-perturbative 
corrections (due to instantons and anti-instantons) can be encoded 
in terms of an elliptic curve, which is very reminiscent to the 
Seiberg-Witten theory. Our four-dimensional results agree with the 
three-dimensional Seiberg-Witten theory and instanton calculations.

\newpage

\section{Introduction}

The exact gauge {\it low-energy effective action} (LEEA) in the
Coulomb branch of N=2 supersymmetric Yang-Mills theory in 
{\it four} spacetime dimensions (4d), which includes both perturbative 
(one-loop) and nonperturbative (instanton) quantum corrections, was 
determined by Seiberg and Witten \cite{sw}. The main tools of their 
construction were the general constraints implied by N=2 extended 
supersymmetry, the known anomaly structure, and electric-magnetic 
duality. The N=2 off-shell supersymmetry implies the unique `Ansatz' 
for the N=2 (abelian) vector multiplet LEEA, in terms of a holomorphic 
function $\cf(W)$ of the N=2 (restricted chiral) superfield strength
$W$. The chiral anomaly determines the perturbative (logarithmic) 
contribution to the function $\cf''(W)$. Nonperturbative consistency 
and duality unambiguously fix the instanton corrections to $\cf(W)$,
which are related to the BPS monopoles representing nonperturbative 
degrees of freedom and belonging to hypermultiplets. The exact 
Seiberg-Witten ($SU(2)$-based) solution can be encoded in terms of an 
elliptic curve $\S_{\rm SW}$, by integrating certain abelian differential 
$\l_{\rm SW}$ over the torus periods.

Since a generic 4d, N=2 gauge field theory has both N=2 vector 
multiplets {\it and} 
hypermultiplets, the latter may also have their own N=2 supersymmetric
LEEA \cite{rev2}. The hypermultiplet LEEA is highly constrained by N=2 
extended supersymmetry and its automorphisms too, so that its exact
form can also be determined. For instance, in three spacetime dimensions, 
Seiberg and Witten \cite{sw3} used the so-called {\it c-map}
\cite{cfg}, relating the special K\"ahler geometry of the N=2 vector 
multiplet moduli space to the hyper-K\"ahler geometry of the 
hypermultiplet moduli space. They further argued that the {\it Atiyah-Hitchin} 
 (AH) metric \cite{ati} is the only regular exact 
solution.~\footnote{An abelian gauge vector is dual to a scalar in
three dimensions.} This proposal was later confirmed by 3d instanton
calculations \cite{dkmtv}, which also discovered a one-parameter family of
possible hyper-K\"ahler metrics generalizing the AH metric. We propose 
the most general `Ansatz' for the 4d hypermultiplet LEEA, which is 
compatible with all unbroken symmetries and has two parameters. It
allows us to reformulate the solution in the very transparent geometrical
way. The earlier approaches \cite{sw3,dkmtv} are formulated only in
3d on the gauge field theory side, and they are not manifestly
supersymmetric, which may cast some doubt on their ultimate consistency.

Our main purpose in this Letter is a derivation of the exact 
hypermultiplet LEEA directly in {\it four} spacetime dimensions, in the
manifestly N=2 supersymmetric way. We confine ourselves  to a {\it
single} hypermultiplet for simplicity. We fully exploit the restrictions
 implied by {\it off-shell} N=2 supersymmetry and its $SU(2)_R$ 
automorphisms, by  making both of them manifest in HSS. Converting 
the HSS action into N=2 {\it projective superspace} (PSS) allows 
us to calculate the effective hyper-K\"ahler metric. The solution can be 
put into the Seiberg-Witten form after introducing the auxiliary elliptic 
curve associated with an $O(4)$ projective multiplet in N=2 PSS.

\section{N=2 supersymmetry and hyper-K\"ahler metrics}

The scalar kinetic part of the hypermultiplet LEEA is of the second
order in spacetime derivatives, so that it has the form of a 
{\it non-linear sigma-model} (NLSM). By N=2 supersymmetry in 4d, 
the metric of this 4d NLSM has to be {\it hyper-K\"ahler}
\cite{hklr}. Making N=2 supersymmetry manifest (i.e off-shell) also
makes manifest the hyper-K\"ahler nature of the hypermultiplet LEEA. 
In the {\it harmonic superspace} (HSS) approach \cite{gikos}, 
both an N=2 vector multiplet and a hypermultiplet 
can be introduced off-shell on equal footing. For example, the 
Fayet-Sohnius hypermultiplet is described by an unconstrained complex 
analytic superfield $q^+$ of $U(1)$ charge $(+1)$, whereas an N=2
vector multiplet is described by an unconstrained analytic superfield 
$V^{++}$ of $U(1)$ charge $(+2)$. 

The general N=2 NLSM Lagrangian in HSS reads \cite{gios}
$$ \cl^{(+4)}= -\sbar{q}{}^+D^{++}q^+ -
\ck^{(+4)}(\sbar{q}{}^{+},q^+;u^{\pm})~,\eqno(2.1)$$
where $\ck^{+(4)}$ is called a hyper-K\"ahler potential, while the harmonic 
covariant derivative $D^{++}$ includes central charges. The N=2
central charge $Z$ can be treated as the abelian N=2 vector superfield 
background whose N=2 gauge superfield strength is given by $\VEV{W}=Z$
\cite{ikz}. The role of the analytic function $\ck^{+(4)}$ in the 
hypermultiplet LEEA (2.1) is similar to the role of the holomorphic 
Seiberg-Witten potential $\cf$ in the N=2 gauge LEEA. Because of 
manifest N=2 supersymmetry by construction, the equations of motion
for the HSS action (2.1) determine (at least, in principle) the
component hyper-K\"ahler NLSM metric in terms of a single analytic 
function $\ck^{(+4)}$. An explicit form of this relation is, however,
not known in general. A crucial simplification arises when the $SU(2)_R$
automorphisms of N=2 supersymmetry are also preserved, which implies that
the hyper-K\"ahler potential is independent upon harmonics. Since  $SU(2)_R$
is known to be non-anomalous \cite{sw,sw3}, the most general `Ansatz' 
for the hypermultiplet LEEA takes the form of a real quartic polynomial, 
$$ \ck^{(+4)}= \fracm{\l}{2}(\sbar{q}{}^+)^2(q^+)^2+\left[ 
\g \sbar{(q^+)}{}^4 +  \b\sbar{(q^+)}{}^3 q^+ +{\rm h.c.}\right]~,\eqno(2.2)$$
with one real $(\l)$ and two complex $(\b,\g)$ parameters. 
The $Sp(1)_{\rm PG}$ internal symmetry of the free hypermultipet action
leaves the form of the quartic (2.2) invariant but not the
coefficients. Hence, $Sp(1)_{\rm PG}$ can be used to reduce the number 
of coupling constants in the family of hyper-K\"ahler metrics
associated with the hyper-K\"ahler potential (2.2) from five to two. 
Equations (2.1) and (2.2) also imply the conservation law \cite{gios} 
$$ D^{++}\ck^{(+4)}=0 \eqno(2.3)$$
on the equations of motion, $D^{++}\sbar{q}{}^+ =\pa\ck^{(+4)}/\pa q^+$
 and  $D^{++}q^+ =-\pa\ck^{(+4)}/\pa\sbar{q}{}^+$.

\section{Perturbative hypermultiplet LEEA}

The manifestly N=2 supersymmetric HSS description of the
hypermultiplet LEEA allows us to exploit the constraints imposed by 
unbroken N=2 supersymmetry and its automorphism  symmetry in the very 
efficient and transparent way. For example, as regards a perturbation 
theory in 4d, N=2 supersymmetric QED (or in the Coulomb branch of N=2
supersymmetric $SU(2)$ Yang-Mills theory \cite{sw}), the unbroken 
symmetry is given by 
$$ SU(2)_{R,~{\rm global}}\times U(1)_{\rm local}~.\eqno(3.1)$$

The {\it unique} hypermultiplet self-interaction consistent with N=2 
supersymmetry and the internal symmetry (3.1) in HSS is described by the 
hyper-K\"ahler potential 
$$ \ck^{(+4)}_{\rm TN}=\fracmm{\l}{2}\left(\sbar{q}{}^+q^+\right)^2~,
\eqno(3.2)$$
just because it is the only function of $U(1)$ charge $(+4)$ that is 
independent upon harmonics and invariant under $U(1)_{\rm local}\,$.  

\begin{figure}
\vglue.1in
\makebox{
\epsfxsize=4in
\epsfbox{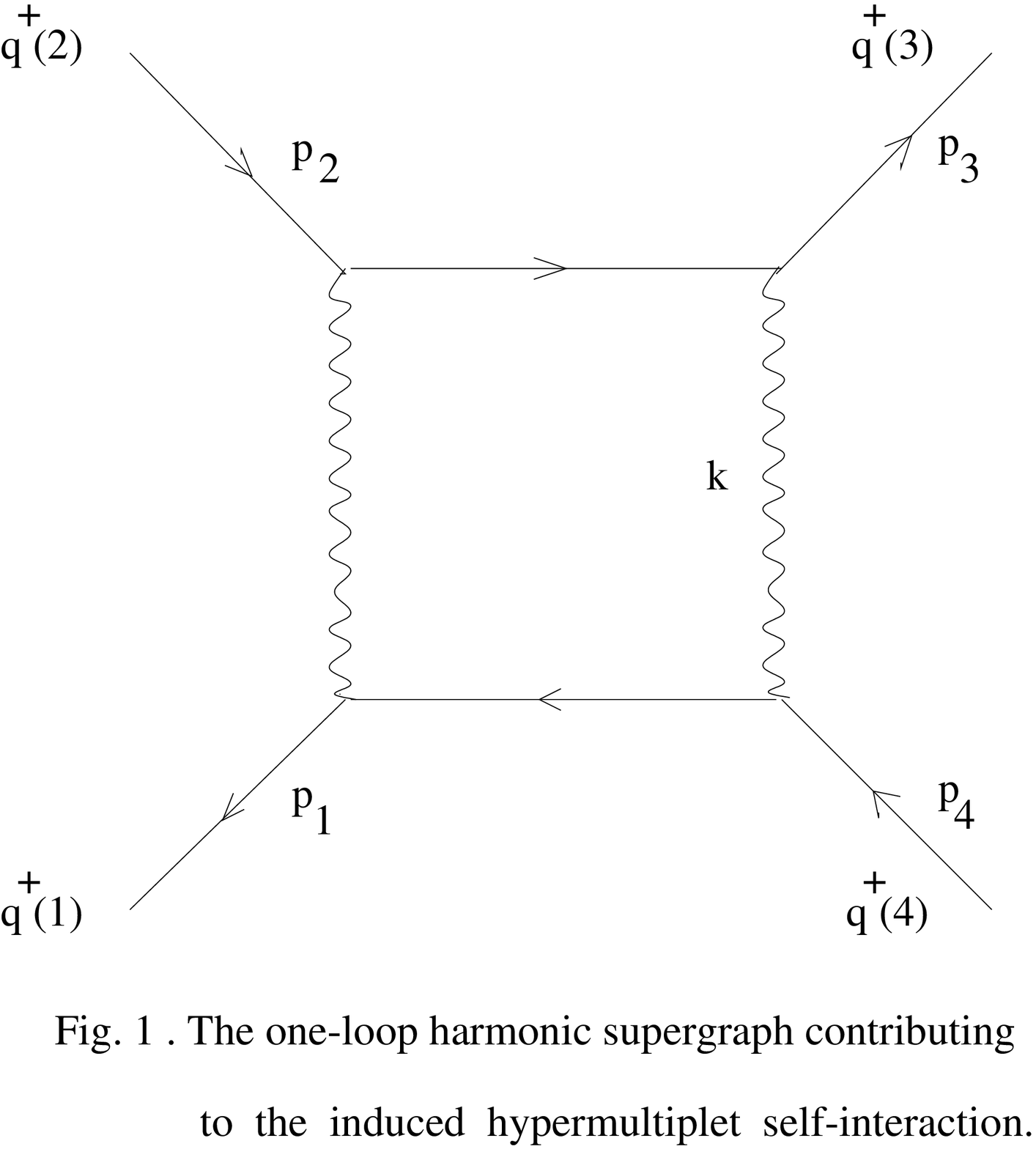}
}
\end{figure}

The coupling constant $\l$ in eq.~(3.2) (in the one-loop
approximation) is determined by the HSS graph shown in Fig.~1.  
The analytic propagator (wave lines in Fig.~1) of the N=2 vector 
superfield (in N=2 Feynman gauge) is (see \cite{ikz} for details) 
$$ i\VEV{ V^{++}(1)V^{++}(2)}=\fracmm{1}{\Box_1}(D_1^+)^4\d^{12}(\cz_1-\cz_2)
\d^{(-2,2)}(u_1,u_2)~.\eqno(3.3)$$
The hypermultiplet analytic propagator (solid lines in Fig.~1) 
with non-vanishing central charges (in the pseudo-real notation)
 reads \cite{ikz}
$$ i\VEV{q^+(1)q^+(2)}=\fracmm{-1}{\Box_1+m^2}
\fracmm{(D^+_1)^4(D^+_2)^4}{(u^+_1u_2^+)^3}e^{\t_3[v(2)-v(1)]}\d^{12}
(\cz_1-\cz_2)~,\eqno(3.4)$$
where $m^2=\abs{Z}^2$ is the hypermultiplet (bare) BPS mass and 
$$ iv =-Z(\bar{\theta}^+\bar{\theta}^-)-\bar{Z}(\theta^+\theta^-)~.\eqno(3.5)$$
In the one-loop approximation one finds \cite{ikz} the predicted form 
(3.2) with 
$$ \l=\fracmm{g^4}{\p^2}\left[ \fracmm{1}{m^2}\ln\left( 1+\fracmm{m^2}{\L^2}
\right) -\fracmm{1}{\L^2+m^2}\right] \eqno(3.6)$$
in terms of the gauge coupling constant $g$, the BPS mass $m^2$ and 
the IR-cutoff $\L$. Note that $\l\neq 0$ only when $Z\neq 0$. The dependence
of $\l$ upon the IR-cutoff is expected to disappear after summing up all
contributions from higher loops.

To understand the hyper-K\"ahler geometry associated with the 
hyper-K\"ahler potential (3.2), it is convenient to rewrite the HSS
action into PSS, by going partially on-shell, in terms of an N=2
{\it tensor} multiplet. Unlike the Fayet-Sohnius 
hypermultiplet, the N=2 tensor superfield $L^{(ij)}$ 
has a finite number of auxiliary fields. In the standard N=2
superspace $(\cz)$  the $L^{(ij)}$ is defined by the off-shell constraints 
$$ D\low{\a}{}^{(i}L^{ik)}=\bar{D}_{\dt{\a}}{}^{(i}L^{jk)}=0~,\eqno(3.7)$$
and the reality condition
$$ \Bar{L^{ij}}=\ve_{ik}\ve_{jl}L^{kl}~.\eqno(3.8)$$
It is not difficult to verify that eq.~(3.7) implies 
$$ \de_{\a}G\equiv (D^1_{\a}+\x D^2_{\a})G=0~,\qquad \D_{\dt{\a}}G\equiv
(\bar{D}^1_{\dt{\a}}+\x\bar{D}^2_{\dt{\a}})G=0~,\eqno(3.9)$$
for {\it any} function $G(Q^2_2(\x),\x)$ depending upon
$$ Q^2_2(\x)\equiv \x_i\x_j L^{ij}(Z)~,\quad \x_i\equiv (1,\x)~, \eqno(3.10)$$
where the $CP(1)$ inhomogeneous coordinate $\x$ has been introduced. 
It follows from eq.~(3.9) that one can construct N=2 invariant actions 
by integrating the potential $G(Q^2_2(\x),\x)$ over the rest of the N=2
superspace coordinates \cite{hklr}, 
$$ S = \int d^4x\fracmm{1}{2\p i}\oint_C \fracmm{d\x}{(1+\x^2)^4}\tilde{\de}^2
\tilde{\D}^2 G(Q^2_2(\x),\x) +{\rm h.c.}~,\eqno(3.11)$$
where the new Grassmann superspace derivatives,
$$ \tilde{\de}_{\a}=\x D^1_{\a}-D^2_{\a}~,\quad \tilde{\D}_{\dt{\a}}=\x
\bar{D}^1_{\dt{\a}}-\bar{D}^2_{\dt{\a}}~,\eqno(3.12)$$
`orthogonal' to those of eq.~(3.9), have been introduced. 
 After being reduced to 4d, N=1 superspace, eqs.~(3.10) and (3.11) 
take the form
$$ \left.Q^2_2(\x)\right| = \F + \x H -\x^2 \bar{\F}~,\eqno(3.13)$$
and
$$ S = \int d^4x d^4\q\fracmm{1}{2\p i}\oint_C \fracmm{d\x}{\x^2}
G(\left.Q^2_2(\x)\right|,\x)+{\rm h.c.}~,\eqno(3.14)$$
in terms of the N=1 chiral superfield $\F$ and the N=1 real linear 
superfield $H$.

The N=2 tensor multipet constraints (3.7) and (3.8) read in HSS as
$$ D^{++}L^{++}=0 \quad {\rm and} \quad \sbar{L}{}^{++}=L^{++}~,\eqno(3.15)$$
respectively, where $L^{++}=u^+_iu^+_jL^{ij}(\cz)$. Let's substitute
(we temporarily set $\l=1$)
$$ \ck^{(+4)}_{\rm TN}=
\fracm{1}{2}(\sbar{q}{}^+q^+)^2=-2(L^{++})^2~,\quad {\rm or,~
equivalently,}\quad \sbar{q}{}^+q^+=2iL^{++}~,\eqno(3.16)$$
which is certainly allowed because of eq.~(2.3). The constraints (3.7) can be 
incorporated off-shell by using extra real analytic superfield $\o$ as 
the Lagrange multiplier. Changing the variables from 
$(\sbar{q}{}^+,q^+)$ to $(L^{++},\o)$ amounts to an N=2 duality 
transformation in HSS. The explicit solution to eq.~(3.16) reads 
$$ q^+=-i\left(2u^+_1+if^{++}u^-_1\right)e^{-i\o/2}~,
\quad \sbar{q}{}^+=i\left( 2u^+_2-if^{++}u^-_2\right)e^{i\o/2}~,\eqno(3.17)$$
where the function $f^{++}$ is given by \cite{gio}
$$ f^{++}(L,u) = \fracmm{2(L^{++}-2iu_1^+u_2^+)}{1+\sqrt{1-4u^+_1u^+_2
u^-_1u^-_2 -2iL^{++}u_1^-u_2^-}}~.\eqno(3.18)$$
It is straightforward to rewrite the free (massless) HSS action
(2.7) in terms of the new variables. It results in the {\it improved} 
(i.e. N=2 superconformally invariant) N=2 tensor multiplet action \cite{gio}
$$S\low{\rm impr.} = \ha \int d\z^{(-4)}du (f^{++})^2~.\eqno(3.19)$$
The action dual to the NLSM action defined by eqs.~(2.1) and (3.2) is 
thus given by a sum of the non-improved (quadratic) and improved 
(non-polynomial) HSS actions for the N=2 tensor multiplet \cite{hklr,gio},
$$ S_{\rm TN}[L;\o]= S\low{\rm impr.} 
+ \ha \int d\z^{(-4)}du \left[(L^{++})^2 +\o D^{++}L^{++}\right]~.\eqno(3.20)$$
The equivalent PSS action is given by eq.~(3.11) with 
$$ \oint G = M\oint_{C_0} \fracmm{(Q^2_2)^2}{2\x} + \oint_{C_r} 
Q^2_2(\ln Q^2_2-1)~,\eqno(3.21)$$
where we have restored the dependence upon $\l$ by setting  $M=\ha\l^{-1/2}$. 
The contour $C_0$ in complex $\x$-plane goes around the origin,
whereas the contour $C_r$ encircles the roots of the quadratic
equation \cite{hklr}
$$ Q^2_2(\x)=0~. \eqno(3.22)$$ 
The hyper-K\"ahler metric of the N=2 NLSM defined by eqs.~(3.21) and (3.22)
is equivalent to the Taub-NUT metric with the mass parameter $M=\ha\l^{-1/2}$
\cite{hklr,gios}.

The $SU(2)_R$ transformations act in PSS in the form of 
{\it projective} (fractional) transformations \cite{oldket} 
$$ \x'=\fracmm{\bar{a}\x-\bar{b}}{a+b\x}~,\qquad 
\abs{a}^2+\abs{b}^2=1~,\eqno(3.23)$$
while a generic PSS action (3.11) is not invariant under these 
transformations. Nevertheless, eq.~(3.11) with some 
non-trivial contour $C_r$ is going to be invariant under the
transformations (3.23) provided that
$$ G({Q_2^2}'(\x'),\x')=\fracmm{1}{(a+b\x)^2}G(Q_2^2(\x),\x)~,\quad 
{\rm where}\quad
{Q_2^2}'(\x')=\fracmm{1}{(a+b\x)^2}Q_2^2(\x)~.\eqno(3.24)$$

Eq.~(3.24) implies that the invariant PSS potential $G(Q^2_2)$ should 
be `almost' linear in $Q^2_2$, like in the second term of the action
 (3.21). The transition $u_i\to\x_i=(1,\x)$ describes the
 holomorphic projection of HSS on PSS, where the analytic superfield 
$L^{++}(\z,u)$ is replaced by the holomorphic (with respect to $\x$) 
section $Q^2_2(L,\x)$ of the line bundle $O(2)$ whose fiber is 
parametrized by constrained superfields. The equation (3.13) defines 
the Riemann sphere in ${\bf C}^2$ parametrized by $(Q_2,\x)$.

\section{Exact hypermultiplet LEEA and $O(4)$ bundle}

In an abelian quantum field theory there are no instantons, so that the
one-loop results of sect.~3 are, in fact, {\it exact} in that case. If,
however, the underlying N=2 gauge field theory has a non-abelian gauge
group whose rank is more than one, one expects nonperturbative contributions 
to the LEEA of a single (magnetically charged) hypermultiplet from instantons
and anti-instantons \cite{hwit}. It may happen, e.g., in the Higgs
branch where the gauge symmetry is completely broken. 

Given the most general $SU(2)_R$-invariant hyper-K\"ahler 
potential (2.2), let's make a substitution
$$ \ck^{(+4)}(q,\sbar{q})\equiv\fracm{\l}{2}(\sbar{q}{}^+)^2(q^+)^2+\left[ 
\g\sbar{(q^+)}{}^4 + \b\sbar{(q^+)}{}^3q^+ +{\rm h.c.}\right]
=L^{++++}(\z,u)~,\eqno(4.1)$$
where the real analytic superfield $L^{++++}$ satisfies the
conservation law (2.3),  
$$ D^{++}L^{++++}=0~.\eqno(4.2)$$
Eq.~(4.2) can be recognized as the {\it off-shell} N=2 (standard) 
superspace constraints
$$ D\low{\a}{}^{(i}L^{jklm)}=\bar{D}_{\dt{\a}}{}^{(i}L^{jklm)}=0~,\eqno(4.3)$$
where $L^{++++}=u^+_iu^+_ju^+_ku^+_lL^{ijkl}(\cz)$, while eq.~(4.1) 
implies the reality condition
$$ \Bar{L^{ijkl}}=\ve_{im}\ve_{jn}\ve_{kp}\ve_{lq}L^{mnpq}~,\eqno(4.4)$$
defining together the $O(4)$ projective multiplet \cite{oldket}. Unlike 
the $O(2)$ tensor supermultiplet (sect.~3), the $O(4)$ supermultiplet 
does not have a conserved vector (or a gauge antisymmetric tensor) 
amongst its field components. 

The N=2 invariant PSS action construction (3.11) in terms of a PSS potential 
$G(Q^2_{4}(\x),\x)$ equally applies to the projective $O(4)$ 
supermultiplets,~\footnote{In fact, it also applies to any projective 
$O(2k)$ multiplets satisfying the off-shell N=2 
superspace \newline ${~~~~~}$ constraints generalizing those of
eqs.~(4.3) and (4.4) with $k>2$ \cite{oldket}.} while $L^{ijkl}$ 
should enter the action via the argument \cite{oldket}
$$ Q^2_{4}(\x)=\x_{i}\x_{j}\x_{k}\x_lL^{ijkl}(\cz)~,\quad \x_i=(1,\x)~.
\eqno(4.5)$$
The N=1 superspace projections of the N=2 superfield (4.5) 
and the N=2 invariant PSS action are given by
$$ \left.Q^2_4(\x)\right|=\F + \x H +\x^2V-\x^3\bar{H} +\x^4\bar{\F}~,
\eqno(4.6)$$
and
$$ S = \int d^4x d^4\q\fracmm{1}{2\p i}\oint_C \fracmm{d\x}{\x^2} G
(\left.Q^2_4(\x)\right|,\x)+{\rm h.c.}~,\eqno(4.7)$$
respectively, in terms of the N=1 chiral superfield $\F$, the N=1
complex linear superfield $H$, and the N=1 general (unconstrained)
real superfield $V$ \cite{hklr,oldket}.

The N=1 superfield $V$ enters the action (4.7) as the Lagrange
multiplier, whose elimination via its `equation of motion' implies 
the algebraic constraint \cite{oldket}
$$ {\rm Re}\,\oint \fracmm{\pa G}{\pa Q^2_4}=0~.\eqno(4.8)$$
Eq.~(4.8) reduces the number of independent N=2 NLSM physical real 
scalars from five to four, which is consistent with the well-known 
fact that the real dimension of any hyper-K\"ahler manifold is a 
multiple of four \cite{hklr}. After solving the constraint (4.8), the complex 
linear N=1 superfield $H$ can be traded for yet another N=1 chiral 
superfield $\J$, by the use of the N=1 superfield Legendre transform 
that results in the N=1 superspace K\"ahler potential 
$K(\F,\J,\bar{\F},\bar{\J})$ associated with the N=2 supersymmetric
 NLSM of eq.~(4.7) \cite{hklr}.

The most straightforward procedure of calculating the dependence $q(L)$, 
as well as performing an explicit N=2 transformation of the 
unconstrained HSS action into the PSS action in terms of the 
constrained N=2 superfield defined by eq.~(4.1), use roots of 
the quartic polynomial. Remarkably, the N=2 PSS action in question 
can be fixed without calculating the roots in the manifestly N=2
supersymmetric approach. It is the $SU(2)_R$ invariance that is powerful
enough to fix the PSS action equivalent to the HSS action of
eqs.~(2.1) and (2.2) ({\it cf.} ref.~\cite{gio}). 
The one real and two complex constants, $(\l,\b,\g)$, respectively, 
parametrizing the hyper-K\"ahler potential (2.2), are naturally united 
into the $SU(2)$ 5-plet $c^{ijkl}$ subject to the reality condition
(4.4). After extracting a constant piece out of $q^+$, 
say, $q^+_a=u^+_a + \tilde{q}^+_a$ and $u_a=(1,\x)$, and collecting
all constant pieces on the left-hand-side of eq.~(4.1), we can
identify their sum with a constant piece
$c^{++++}=c^{ijkl}u^+_iu^+_ju^+_ku^+_l$ of $L^{++++}$ on the right-hand-side
of eq.~(4.1), representing the constant vacuum expectation values of
the N=1 superfield 
components of $L^{++++}$ defined by eq.~(4.6), i.e.
$$ \l=\VEV{V}~,\quad \b=\VEV{H}~,\quad\g=\VEV{\F}~.\eqno(4.9)$$

The $SU(2)_R$ transformations in PSS are the projective transformations 
(3.24), so that the PSS potential $G$ of the `improved' $O(4)$
multiplet action having the form (3.11) must be proportional to
 $Q_4\equiv\sqrt{Q^2_4}$  because of the relations  
$$G({Q_4^2}'(\x'),\x')=\fracmm{1}{(a+b\x)^2}G(Q_4^2(\x),\x) \quad{\rm and}\quad
 {Q^2_4}'(\x')=\fracmm{1}{(a+b\x)^4}Q^2_4(\x)~.\eqno(4.10)$$
The most general non-trivial contour $C_r$ in complex $\x$-plane, 
whose definition is compatible with the projective $SU(2)$ symmetry, 
is the one encircling the roots of the quartic ({\it cf.} sect.~3),
$$\left.Q^2_4(\x)\right|=p + \x q + \x^2 r -\x^3\bar{q}+\x^4\bar{p}~,
\eqno(4.11)$$ with one real $(r)$ and two complex $(p,q)$ 
additional parameters belonging to yet another 5-plet of $SU(2)$. 
The projective $SU(2)$ invariance of the PSS action defined by
eqs.~(4.7) and (4.11) can be used to reduce the number of independent 
parameters in the corresponding family of hyper-K\"ahler metrics 
from five to two, which is consistent with the HSS predictions of 
sect.~2.~\footnote{The generalization of eq.~(3.22) similarly to eq.~(4.11) 
is `empty' since the quadratic polynomial 
\newline ${~~~~~}$ $c^2_2(\x)=p+\x r-\x^2\bar{p}$ can
always be removed by an $SU(2)$ transformation.} We didn't attempt 
to establish an explicit relation between the HSS coefficients 
$(\l,\g,\b)$ and the PSS coefficients $(r,q,p)$. The most natural (non-trivial)
contour $C_r$ surrounds the roots of the equation
$$ Q^2_4(\x)=0~, \eqno(4.12)$$
and it leads to the only non-singular hyper-K\"ahler NLSM metric (sect.~5).

The $SU(2)$-invariant PSS action, equivalent to the one defined by eqs.~(2.1)
and (4.1), is therefore given by
$$ \fracmm{1}{2\p i}\oint G =-\fracmm{1}{2\p i}\oint_{C_0} \fracmm{Q^2_4}{\x} +
\oint_{C_r} Q_4~.\eqno(4.13)$$
The constraint (4.8) in the case (4.12) takes the form 
$$ \oint_{C_r} \fracmm{d\x}{\sqrt{Q^2_4}}=1~.\eqno(4.14)$$

The component form of the metric associated with eqs.~(4.13) and (4.14) 
was found in ref.~\cite{iro}. Because of the reality condition (4.4),
 the quartic (4.12) has two pairs of roots $(\r,-1/\bar{\r})$ related by 
an $SL(2,{\bf Z})$ transformation and satisfying the defining relation
$$ Q^2_4(\x)=c(\x-\r_1)(\bar{\r}_1\x+1)(\x-\r_2)(\bar{\r}_2\x+1)~.\eqno(4.15)$$
The branch cuts of the root in eq.~(4.14) can be chosen to run from
$\r_1$ to $-1/\bar{\r}_2$ and from $\r_2$ to $-1/\bar{\r}_1$. The
contour integration in eq.~(4.14) can thus be reduced to the complete
elliptic integral (in the Legendre normal form) over the branch cut \cite{iro},
$$ \fracmm{4}{\sqrt{c(1+\abs{\r_1}^2)(1+\abs{\r_2}^2)}}
\int^1_0\fracmm{d\x}{\sqrt{(1-\x^2)(1-k^2\x^2)}}=1~,\eqno(4.16)$$
with the modulus
$$ k^2= \fracmm{(1+\r_1\bar{\r}_2)(1+\r_2\bar{\r}_1)}{(1+\abs{\r_1}^2)
(1+\abs{\r_2}^2)}~.\eqno(4.17)$$
The constraint (4.16) can be explicitly solved in terms of the
complete elliptic integrals,
$$ K(k)=\int^{\p/2}_0 \,\fracmm{d\g}{\sqrt{1-k^2\sin^2\g}} \quad
{\rm and} \quad
E(k) =\int^{\p/2}_0 d\g\,\sqrt{1-k^2\sin^2\g}~~,\eqno(4.18)$$
of the first and second kind, respectively, by using the following 
parametrization \cite{iro}:
$$ \eqalign{ 
\F ~=~&  2e^{2i\vf} \left[ \cos(2\j) (1+\cos^2\vq)\right. \cr 
& \left.+ 2i\sin(2\j)\cos\vq + (2k^2-1)\sin^2\vq \right] K^2(k)~,\cr
H ~=~&  8e^{i\vf}\sin\vq \left[ \sin(2\j) \right. \cr
& \left.- i\cos(2\j)\cos\vq + i(2k^2-1)\cos\vq \right] K^2(k)~,\cr
V ~=~&  4\left[ -3\cos(2\j)\sin^2\vq +(2k^2-1)(1-3\cos^2\vq)\right]
K^2(k)~,\cr}\eqno(4.19)$$
in terms of the Euler `angles' $(\vq,\j,\vf)$ and the modulus $k$ 
representing the independent (superfield) coordinates of the N=2 NLSM 
under consideration. Applying the generalized Legendre transform \cite{iro}
to the function (4.13) with respect to $H$ gives rise to the 
{\it Atiyah-Hitchin} (AH) metric \cite{ati}
$$ ds^2_{\rm AH}=\fracmm{1}{4}A^2B^2C^2\left(\fracmm{dk}{k{k'}^2K^2}\right)^2
+ A^2(k)\s_1^2 + B^2(k)\s^2_2 + C^2(k)\s^2_3~,\eqno(4.20)$$
whose coefficient functions satisfy the relations \cite{ati}
$$ \eqalign{
AB ~=~ & -K(k)\left[E(k)-K(k)\right]~,\cr
BC~=~ & -K(k)\left[E(k)-{k'}^2K(k)\right]~,\cr
AC~=~ & -K(k)E(k)~.\cr}\eqno(4.21)$$
while $\s_i$ stand for the $SO(3)$-invariant one-forms
$$ \eqalign{
\s_1 ~=~ & +\frac{1}{2}\left(\sin\j d\vq - \sin\vq \cos\j d \vf\right)~,\cr
\s_2 ~=~ & -\frac{1}{2}\left(\cos\j d\vq + \sin\vq \sin\j d \vf\right)~,\cr
\s_3 ~=~ & +\frac{1}{2}\left(d\j + \cos\vq d\vf\right)~,\cr}\eqno(4.22)$$
and $k'$ is known as the complementary modulus, ${k'}^2=1-k^2$.

In the limit $k\to 1$ (or, equivalently, $k'\to 0$), one has an 
asymptotic expansion
$$K(k)\approx -\log\, k'\left[ 1 +\fracmm{(k')^2}{4}\right] +\ldots
\eqno(4.23)$$
Eq.~(4.23) suggests us to make a redefinition 
$$ k'=\sqrt{1-k^2}\approx 4\exp\left(\fracmm{1}{\g}\right)~,\eqno(4.24)$$
and describe the same limit at $\g\to 0^-$. After substituting eq.~(4.23) 
into eq.~(4.21) one finds that the AH metric becomes 
{\it exponentially close} to the Taub-NUT metric in
the form (4.20) subject to the additional relations: 
$$ A^2\approx B^2\approx \fracmm{1+\g}{\g^2}~~,\quad C^2\approx 
\fracmm{1}{1+\g}~~.\eqno(4.25)$$
The extra $U(1)$ symmetry of the Taub-NUT metric is the direct 
consequence of the relation $A^2=B^2$ arising from the AH metric in the 
asymptotic limit described by eq.~(4.25). The vicinity of $k'\approx
0^+$ describes the region of the hypermultiplet moduli space where 
quantum perturbation theory applies, with the exponentially small AH 
corrections to the Taub-NUT metric being interpreted as the
one-instanton and anti-instanton contributions to the hypermultiplet
LEEA \cite{sw3,dkmtv}.

From the N=2 PSS viewpoint, the transition from the perturbative 
hypermultiplet LEEA to the nonperturbative one corresponds to the 
transition from the $O(2)$ holomorphic line bundle associated with the 
standard N=2 tensor supermultiplet to the $O(4)$ holomorphic line bundle 
associated with the $O(4)$ N=2 supermultiplet. The two holomorphic
bundles are topologically different: with respect to the standard 
covering of $CP(1)$ by two open affine sets, the $O(2)$ bundle has 
transition functions $\x^{-1}$, whereas the $O(4)$ bundle has
transition functions  $\x^{-2}$. The variable $Q$ is the coordinate of 
the corresponding fiber over $CP(1)$. 

\section{Atiyah-Hitchin metric and elliptic curve}

The quadratic dependence of $Q^2_2$ on $\x$ in eqs.~(3.10) and (3.13) 
allows us to globally 
interpret it as a holomorphic (of degree 2) section of PSS, fibered by
the superfields $(\F,H)$ and topologically equivalent to a complex
line (or Riemann sphere of genus $0$). Similarly, the 
quartic dependence of $Q^2_4$ on $\x$ in eqs.~(4.5) and (4.6) allows
us to globally interpret it as a holomorphic (of degree 4) section of 
PSS, fibered by the superfields $(\F,H,V)$ and topologically
equivalent to an {\it elliptic curve} $\S_{\rm hyper.}$ (or a torus 
of genus $1$). The non-perturbative hypermultiplet LEEA can 
therefore be encoded in terms of the genus-one 
Riemann surface $\S_{\rm hyper.}$ in close analogy to the exact 
N=2 gauge LEEA in terms of the elliptic curve $\S_{\rm SW}$ 
of Seiberg and Witten \cite{sw}.

The classical twistor construction of hyper-K\"ahler metrics 
\cite{ati} is known to be closely related to the {\it Hurtubise} 
elliptic curve $\S_{\rm H}$ \cite{hur}. This curve can be identified 
with $\S_{\rm hyper.}$ that carries the same information and
whose defining eq.~(4.6) can be put into the Hurtubise form, 
$$ \tilde{Q}^2_4(\tilde{\x})= 
K^2(k)\tilde{\x}\left[ kk'(\tilde{\x}^2-1)+(k^2-{k'}^2)\tilde{\x}
\right]~,\eqno(5.1)$$
by a projective $SU(2)$ transformation. In its turn, eq.~(5.1) is
simply related (by a linear transformation) to another standard 
(Weierstrass) form, $y^2=4x^3-g_2x-g_3$. Therefore, in accordance with 
ref.~\cite{ati}, the real period $\o$ of $\S_{\rm H}$ is
$$\o\equiv 4k_1~,\quad {\rm where}\quad 4k_1^2=kk'K^2(k)~,\eqno(5.2)$$
whereas the complex period matrix of $\S_{\rm H}$ is given by
$$\t=\fracmm{iK(k')}{K(k)}~.\eqno(5.3)$$

At generic values of the AH modulus $k$, $0<k<1$, the roots of the 
Weierstrass form are all 
different from each other, while they all lie on the real axis, say, at 
$e_3<e_2<e_1<\infty=(e_4)$. Accordingly, the branch cuts are running 
from $e_3$ to $e_2$ and from $e_1$ to $\infty$. The $C_r$ integration 
contour in the PSS formulation of the exact 
hypermultiplet LEEA in eq.~(4.13) can now be interpreted as the 
contour integral {\it over the non-contractible $\a$-cycle of the 
elliptic curve} $\S_{\rm H}$ \cite{bakas}, again in the very similar 
way as the Seiberg-Witten solution to the $SU(2)$-based N=2 gauge LEEA 
is written down in terms of the abelian differential $\l_{\rm SW}$ 
integrated over the periods of $\S_{\rm SW}$ \cite{sw}. The most
general (non-trivial) integration contour $C_r$ in eq.~(4.13) 
is given by a linear combination of the non-contractible $\a$ and $\b$ 
cycles of $\S_{\rm H}$, while an integration over $\b$ is known to lead 
to a singularity \cite{ati}. This simple observation implies that the 
AH metric is the only regular solution.

The perturbative (Taub-NUT) limit $k\to 1$ corresponds to the
situation when $e_2\to e_1$, so that the $\b$-cycle of $\S_{\rm H}$ 
degenerates. The curve (5.1) then asymptotically approaches a complex line, 
$\tilde{Q}_4\sim \pm K\tilde{\x}$. Another limit, $k\to 0$, leads to a 
(coordinate) bolt-type singularity of the AH metric in the standard 
parameterization (4.20) \cite{ati}. In the context of monopole
physics, this corresponds to the coincidence limit of 
two centered monopoles. In the context of the hypermultiplet LEEA, 
$k\to 0$ implies $e_2\to e_3$, so that the $\a$-cycle of $\S_{\rm H}$ 
degenerates, as well as the whole hypermultiplet 
action associated with eq.~(4.13). The two limits, $k\to 1$ and 
$k\to 0$, are related by the modular transformation exchanging $k$ 
with $k'$, and $\a$-cycle with $\b$-cycle \cite{bakas}. The 
non-perturbative corrections to the hypermultiplet LEEA are 
therefore dictated by the hidden (in 4d) elliptic curve parametrizing 
the exact solution. 

\section*{Acknowledgements}

I am grateful to Ioannis Bakas, Jim Gates and Olaf Lechtenfeld 
for useful discussions.

\end{document}